# Betatron modulation of microwave instability in small isochronous ring

Yingjie Li*
Department of Physics, Michigan State University, East Lansing, MI 48824, USA

Lanfa Wang
SLAC National Accelerator Laboratory, Menlo Park, CA 94025, USA

A novel coupling of the transverse betatron motion to the longitudinal microwave instability is studied. Besides the radial coherent dipole mode space charge field, simulation and theoretical studies in this paper show that the longitudinal coherent dipole mode space charge field due to centroid wiggles also plays an important role in the isochronous regime, it induces betatron oscillation frequencies in temporal evolutions of spectra of longitudinal charge densities, radial centroid offsets and coherent energy deviations of local centroids.

## I. INTRODUCTION

In recent years, high power isochronous cyclotrons have been considered for applications in scientific research, medical therapy, etc. A bottleneck that limits the operation of a high power isochronous cyclotron is the beam instability induced by the space charge force. Gordon [1] explained the physical origin of vortex motion and deformation of beam shape in cyclotrons. In the last decade, additional extensive studies on the space charge effects in isochronous regime have been done through numerical simulations, experiments and analytical models [2-9].

Pozdeyev and Bi proposed their own models and theories to explain the mechanisms of microwave instability of a coasting long bunch with space charge in circular accelerators operating in the isochronous regime [5-7], respectively. The two models only take into account the radial coherent dipole mode space charge field due to beam centroid wiggles and longitudinal monopole mode space charge field originating from line charge density modulations, none of them discusses the effects of longitudinal coherent dipole mode space charge field on evolution of beam instability due to beam centroid wiggles. The perturbation wavenumber k in Pozdeyev's model and growth rate formula [5, 6] is defined in the sinusoidal radial centroid offset function of a beam with uniform charge density, while in Bi's model and growth rate formula [7], the perturbation wavenumber k is defined in the line density modulation function. Actually the growth rates spectra of the line charge densities and radial centroid offsets are different. The relation and interaction between the spectral evolutions of two parameters of different meanings in physics are not discussed in the two models. The missing term of unperturbed line density in the calculation of radial coherent space charge field in [7] also makes the growth rates formula not compatible with the scaling law with respect to beam intensity observed in simulations and experiments [3, 5, 6].

Simulation studies in this paper show that the microwave instability in isochronous regime is usually characterized by betatron oscillations superimposed on exponential growths. These phenomena cannot be explained by the two existing models with conventional 1-dimensional microwave instability growth rate formula which can only predict pure exponential growths. A theoretical discussion in this paper explains that the above beam behaviors primarily originate from the longitudinal coherent dipole mode space charge field of the beam centroid wiggles, the interaction and correlation of temporal spectral evolutions between line charge densities, radial offsets and energy deviations of local centroids are also revealed by a set of longitudinal and radial equations of motion taking into account both the radial and longitudinal coherent dipole mode space charge fields.

This paper is organized as follows. Sec. II gives a brief introduction to Small Isochronous Ring (SIR) and simulation code used. Simulation and theoretical studies of the temporal spectral evolution of beam parameters affected by both longitudinal and radial coherent space charge fields are provided in Sec. III and IV, respectively.

---

* liyingji@msu.edu

## II. SMALL ISOCHRONOUS RING AND CYCO

To simulate and study beam dynamics, especially the space charge effects in high power isochronous cyclotrons, a low energy, low beam intensity Small Isochronous Ring (SIR) was constructed at the National Superconducting Cyclotron Lab (NSCL) at Michigan State University (MSU) [3, 4]. Its main parameters are shown in Table 1.

Table 1 Main Parameters of SIR

| Parameters | Values |
|---|---|
| Ion species | $H_2^+$ |
| Kinetic energy | 20 keV |
| Beam current | 5-25 $\mu$A |
| Bunch length | 15 cm- 5.5 m |
| Betatron tunes | $\nu_x$=1.14, $\nu_y$=1.11 |
| Slip factor | $\eta_0$= 2×10$^{-4}$ |
| Circumference | 6.58 m |
| Rev. period | 4.77 $\mu$s |
| Life time | ~ 200 turns |
| Beam radius | ~ 0.5 cm |
| Chamber width | 11.4 cm |
| Chamber height | 4.8 cm |

The Small Isochronous Ring consists of a multi-cusp Hydrogen ion source, an injection line and a storage ring. The ion source can produce three species of Hydrogen ions, an analyzing dipole magnet under the ion source is used to select the $H_2^+$ ions which are usually used in the experiments. The $H_2^+$ ion beam with desired bunch length can be produced by a chopper and its Courant-Snyder parameters may be matched to the storage ring by an electrostatic quadruple triplet. The storage ring mainly consists of four identical flat-field bending magnets with edge focusing; the 26$^o$ rotation angle of pole face of each magnet provides both vertical focusing and isochronism. There is no RF cavity in the storage ring. After injection to the storage ring by a pair of fast-pulsed electrostatic inflectors, the bunch can coast in the ring up to 200 turns. There is an extraction box located in the drift line between the 2$^{nd}$ and 3$^{rd}$ bending magnets, a pair of fast pulsed electrostatic deflectors in the extraction box can kick the beam either up to a phosphor screen above the median ring plane, or down to the Fast Faraday Cup (FFC) below the median ring plane. The phosphor screen and FFC are used to monitor the transverse and longitudinal beam profiles, respectively. We can also perform energy spread measurement if the FFC assembly is replaced by an energy analyzer assembly.

CYCO [3] is a 3-dimensional Particle-In-Cell (PIC) simulation code that was developed by Pozdeyev to study the beam dynamics with space charge in isochronous regime. It can numerically solve the complete and self-consistent system of six equations of motion of charged particles in a realistic 3D field map including space charge field. Because of large aspect ratio between the width and height of the vacuum chamber of storage ring, the code only includes the image charge effects in the vertical direction. The rectangular vacuum chamber is simplified as a pair of infinitely large ideally conducting plates parallel to the median ring plane.

## III. SIMULATION STUDIES ON EVOLUTIONS OF BEAM PARAMETERS IN ISOCHRONOUS RING

In order to acquire detailed information and properties of microwave instability in isochronous ring, in this section, we will present the simulation methods and results for long-term spectral evolutions of beam parameters based on Fast Fourier Transform (FFT) technique. Here the term 'long-term' denotes a time scale of multiple betatron oscillation periods.

The simulation study of linear stage of long-term microwave instability was carried out for a macro-particle bunch with initial beam intensity $I_0 = 10$ uA, kinetic Energy $E_{k0} = 19.9$ keV, bunch length $\tau_b = 300$ ns (~40 cm), radial and vertical emittance $\varepsilon_{x,0} = \varepsilon_{y,0} = 50\ \pi$ mm*mrad using CYCO. The bunch has a uniform initial distribution in both 4-dimensional transverse phase space (x, x', y, y') and longitudinal charge density. Figure 1 shows the evolution of top views of beam profiles. The beam moves from left to right. We can observe the beam shape is deformed and amplitudes of line charge density modulations increase with turn number due to space charge force.

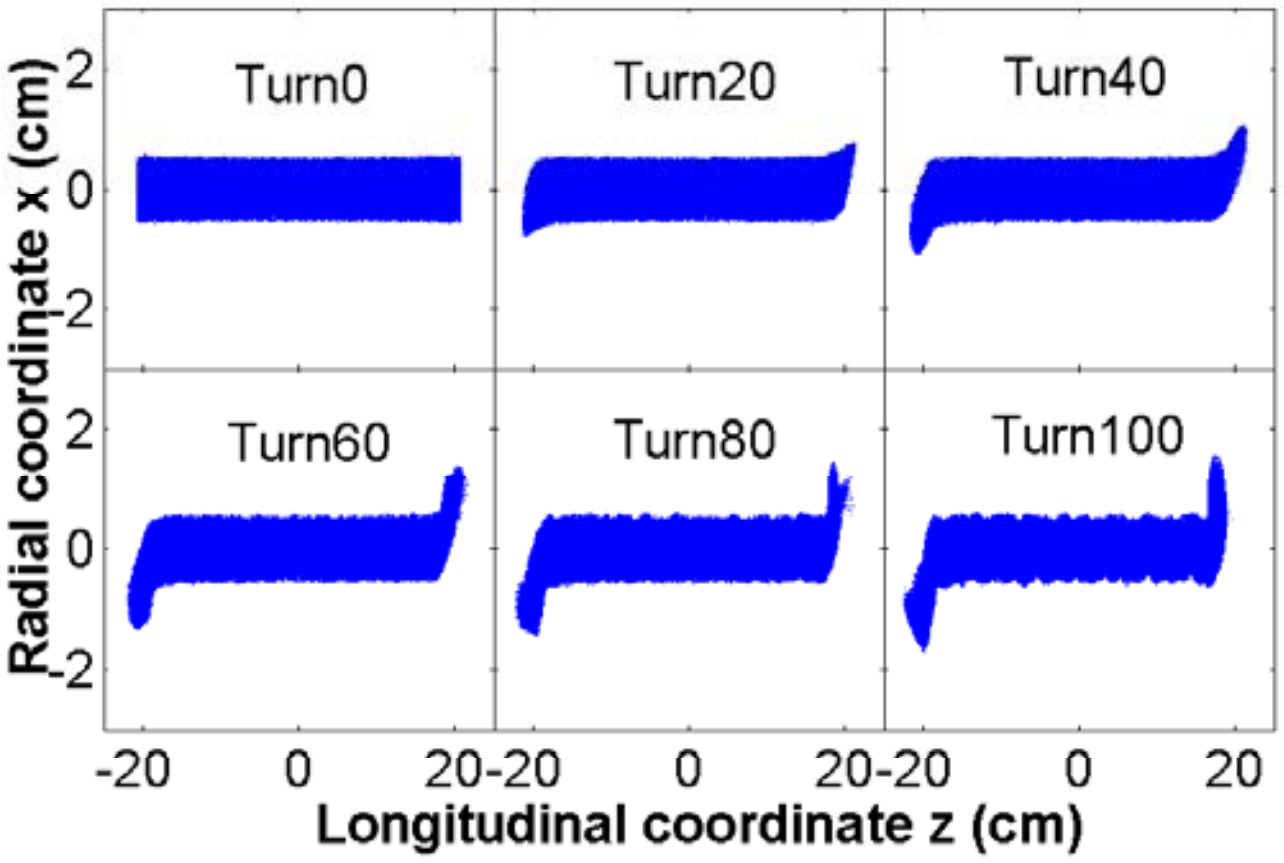

FIG. 1. (Color) Evolution of top views of beam profiles.

FFT analysis was performed for spectral evolutions of line charge densities, radial offsets and energy deviations of centroids with respect to longitudinal coordinate z. Considering the strong nonlinear beam dynamics in bunch head and tail, only the beam profiles of central half of the bunch is used in FFT analysis. The analysis results of some chosen perturbation wavelengths are shown in Figure 2 - Figure 4, respectively.

It is clearly shown that there are many oscillations superimposed on the exponential growth curves in these figures. Because the radial betatron tune of SIR beam is 1.14, the betatron oscillation period is about $1/0.14 \approx 7$ turns, it is easy to judge that these oscillations are induced by the coupling of betatron oscillations. Figure 2 shows coupling from transverse betatron oscillations to longitudinal line charge densities. It is the first time to clearly observe this coupling to our knowledge. Actually, an indication of similar oscillations can be found in Figure 9 of [6], where there is a fast instability growth due to high beam intensity.

The betatron oscillations in Figure 2 cannot be explained by the existing models and theories of microwave instability in [6-7], which can only predict pure exponential growth of perturbed line charge densities. In fact, they are induced by the longitudinal coherent dipole mode space charge field due to centroid wiggles which will be explained in Sec. IV.

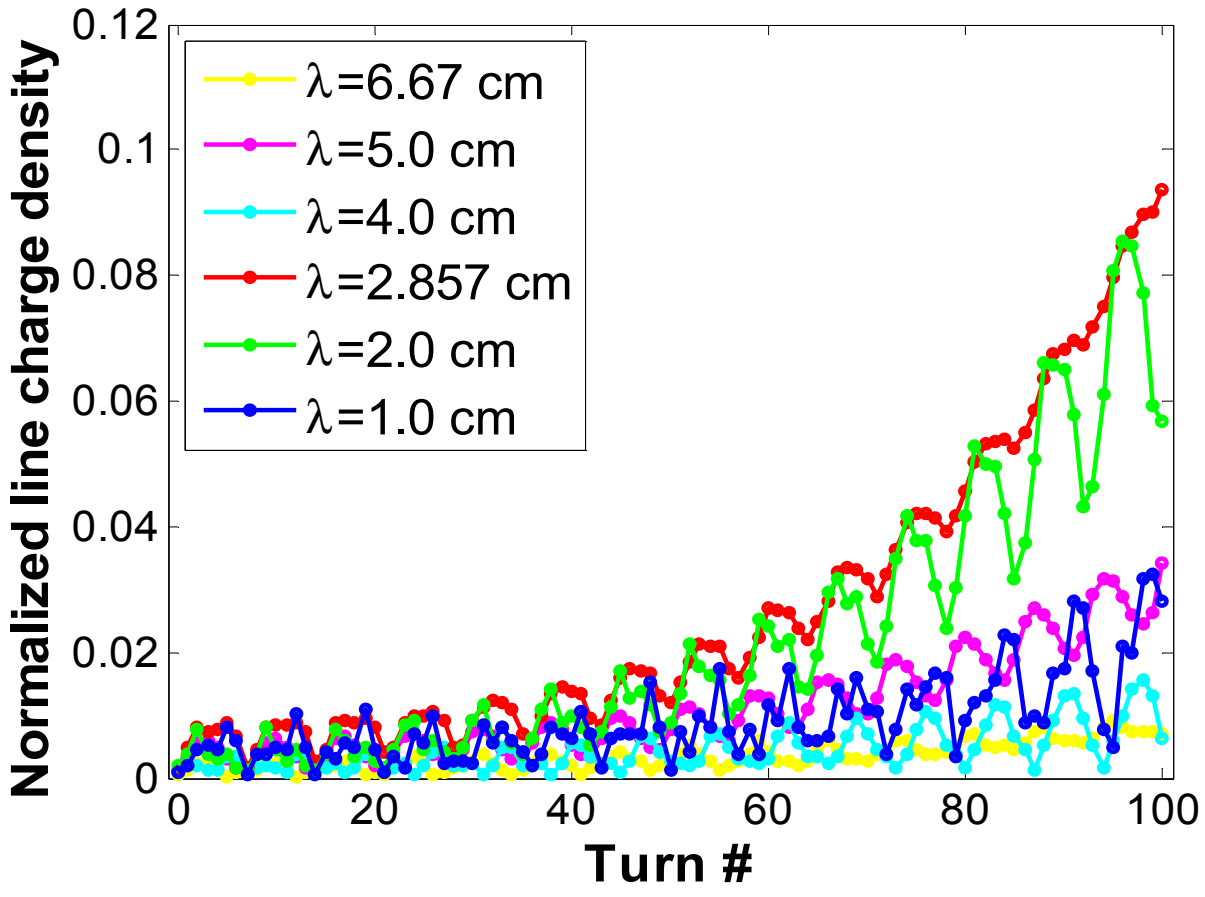

FIG. 2. (Color) Evolution of harmonic amplitudes of normalized line charge densities.

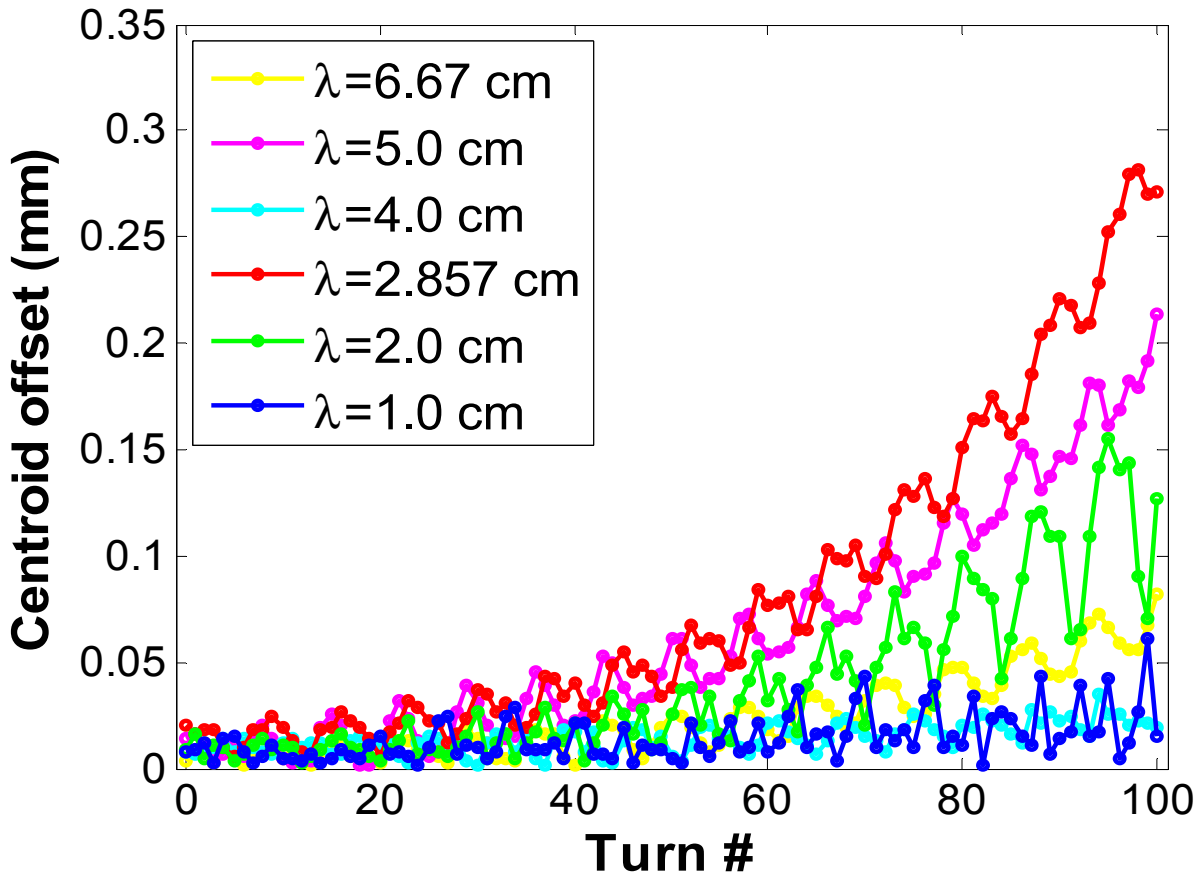

FIG. 3. (Color) Evolution of harmonic amplitudes of radial centroid offsets.

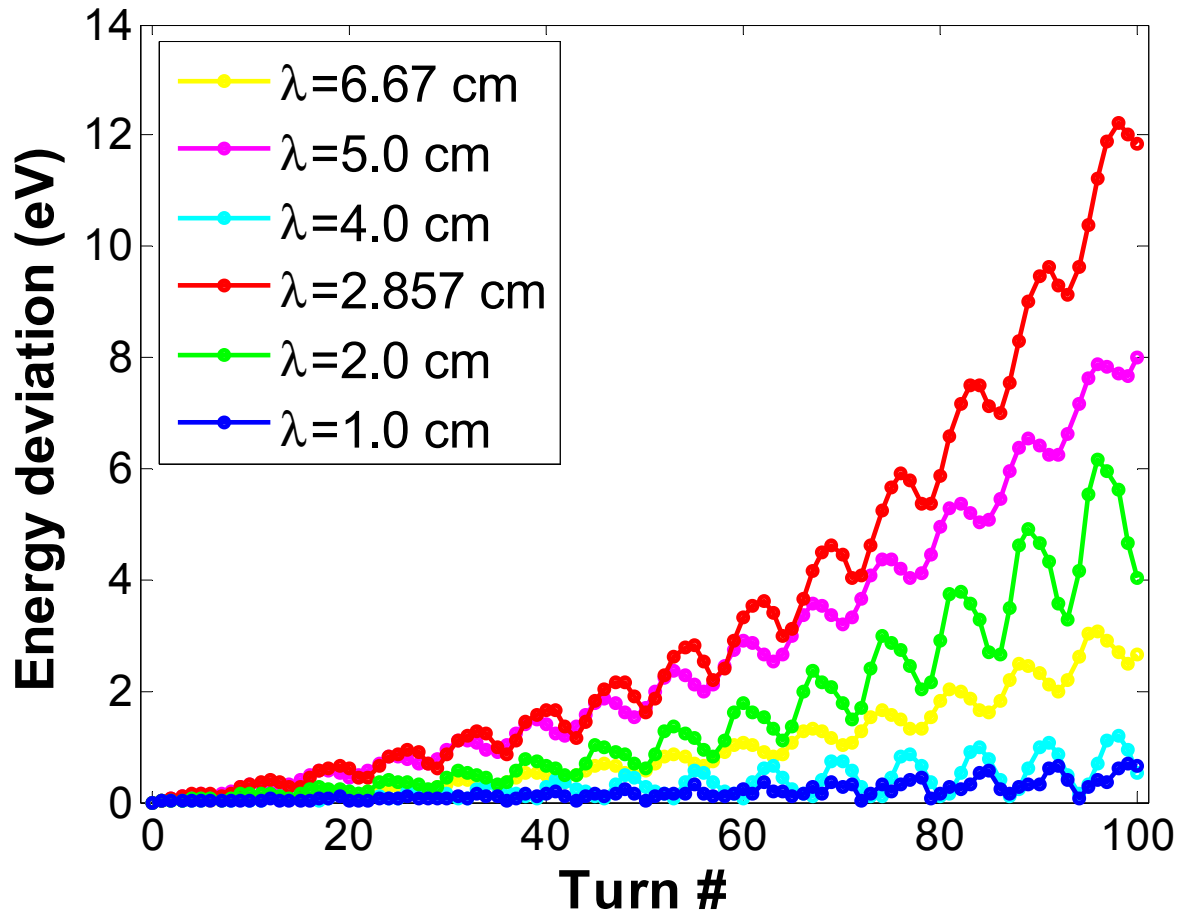

FIG. 4. (Color) Evolution of harmonic amplitudes of energy deviations.

# IV. THEORETICAL STUDIES OF EVOLUTIONS OF BEAM PARAMETERS IN ISOCHRONOUS RING

In isochronous ring, the gradient of longitudinal line charge density of a coasting bunch may induce longitudinal space charge fields, coherent energy deviations and the associated centroid offsets. For simplicity, we assume both centroid offset $x_c(z, t)$ and line charge density $\Lambda(z, t)$ consist of only a single harmonic component neglecting the nonlinear coupling between the chosen components and other components. For general purpose, we also assume there is no correlation between wavenumbers $k_c$ of $x_c(z, t)$ and k of $\Lambda(z, t)$. Using the relation of phase $\Phi = ks - \omega t + \phi_0 = k(z + v_0 t) - \omega t + \phi_0 = kz - \omega t + kv_0 t + \phi_0$, where $\phi_0$ is initial phase at t = 0 and s = 0, $v_0 = \beta c$ is the velocity of on-momentum particles, then the local beam centroid $x_c(z, t)$, line charge density $\Lambda(z, t)$ and beam intensity $I(z, t)$ can be expressed as:

$$x_c(z,t) = \hat{a}_c e^{i(k_c z - \omega_c t + \phi_c)} \tag{1}$$

$$\Lambda(z,t) = \Lambda_0 + \Lambda_1(z,t) = \Lambda_0 + \hat{\Lambda} e^{i(kz - \omega t + \phi)} \tag{2}$$

$$I(z,t) = I_0 + I_1(z,t) = I_0 + \hat{I} e^{i(kz - \omega t + \phi)} \tag{3}$$

where $\phi_c(t) = k_c v_0 t + \phi_{0,c}$, $\phi(t) = kv_0 t + \phi_0$, $\hat{I} = \hat{\Lambda}\beta c$, and amplitudes $\hat{a}_c$, $\hat{\Lambda}$, $\hat{I}$ are all real numbers. $\omega$ and $\omega_c$ are perturbation frequencies of line charge densities and radial centroid offsets, respectively.

## A. Longitudinal coherent dipole mode space charge field and impedance

The centroid wiggles produce not only the radial coherent dipole mode space charge field $E_x$ as pointed out by [6], but also its longitudinal counterpart $E_s^{(1)}$ as shown in Figure 5.

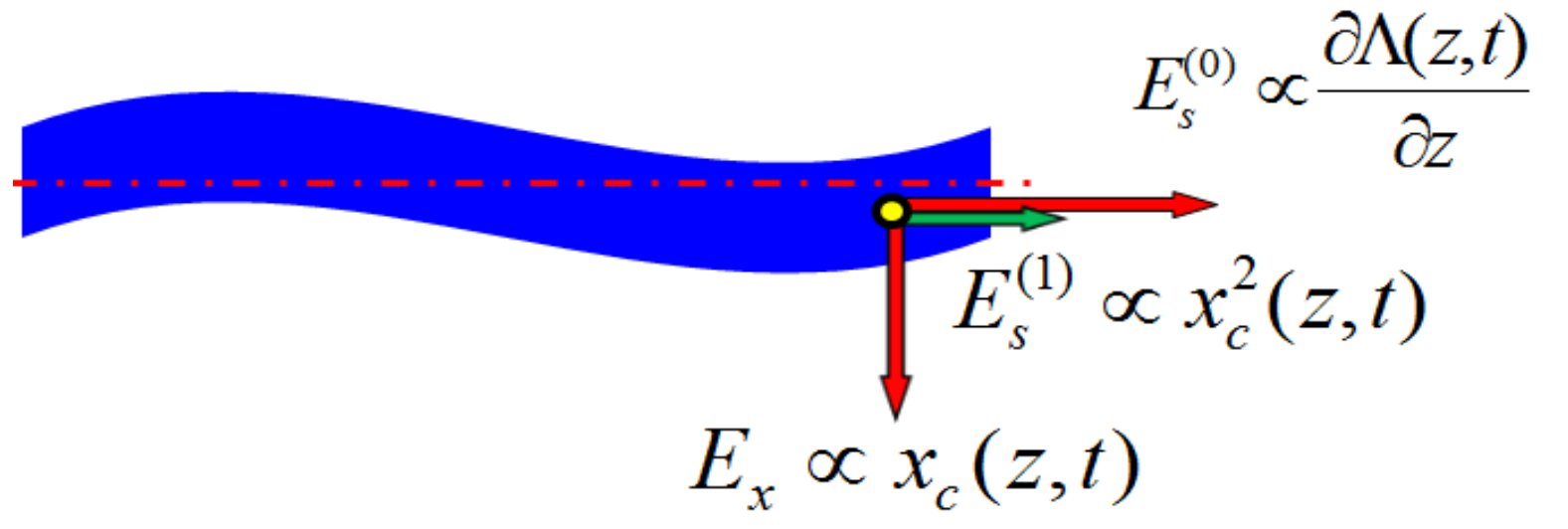

FIG. 5. (Color) Space charge components on local centroid with coordinate $(x_c, z)$ at time t.

The total longitudinal coherent space charge field on a local beam centroid can be approximated as:

$$E_s(z,t) = E_s^{(0)}(z,t) + E_s^{(1)}(z,t) \tag{4}$$

where the first and second terms in Eq. (4) are longitudinal coherent space charge fields of monopole and dipole modes generated by line charge density modulations and centroid wiggles, respectively. If we adopt Pozdeyev's circular beam model of radius $r_0$ in free space [6], the first term of Eq. (4) can be approximated by on-axis value of space charge field of circular beam in free space, which can be calculated by setting r = 0 and radius of chamber wall $r_w = \infty$ in Eq. (17) of [7] as:

$$E_s^{(0)}(z,t) = -i\frac{\hat{\Lambda}e^{i(kz-\omega t+\phi)}}{\pi\varepsilon_0 r_0^2 k}[1-kr_0K_1(kr_0)] \tag{5}$$

Where $\varepsilon_0 = 8.85\times10^{-12}$ F m$^{-1}$ is the permittivity of free space, $r_0$ is beam radius, $k$ is line charge density perturbation wavenumber, $K_1(x)$ is the modified Bessel function of the second kind.

The corresponding longitudinal monopole mode space charge impedance in low energy and short wavelength limit is [6]:

$$Z(k) = Z_{0,sc}^{\parallel}(k) = i\frac{2Z_0R}{k\beta r_0^2}[1-kr_0K_1(kr_0)] \tag{6}$$

Where $Z_0 = 377\ \Omega$ is the impedance in free space, $R$ is average ring radius, β is relativistic speed factor.

When a beam has centroid wiggles induced by longitudinal space charge field, according to Eq. (4) of [6], the radial space charge field on a particle with coordinate (x, z) can be estimated in SI unit system as:

$$E_x(x,z,t) = \frac{\Lambda_0}{2\pi\varepsilon_0 r_0^2}[x - x_c(z,t)k_cr_0K_1(k_cr_0)] \tag{7}$$

Where $\Lambda_0$ is the unperturbed part of line charge density, $x_c(z, t)$ is the time-dependent local radial centroid offset.
Let $x = x_c(z, t)$ in Eq. (7), the radial coherent dipole mode space charge field on local centroid becomes:

$$E_x(z,t) = \frac{\Lambda_0}{2\pi\varepsilon_0 r_0^2}[1 - k_cr_0K_1(k_cr_0)]x_c(z,t) \tag{8}$$

Because the curl of electric field is 0, the longitudinal coherent dipole mode space charge field on a particle can be calculated as:

$$E_s^{(1)}(x,z,t) = x\frac{\partial E_s(x,z,t)}{\partial x} = x\frac{\partial E_x(x,z,t)}{\partial z} = -\frac{\Lambda_0}{2\pi\varepsilon_0 r_0}k_cK_1(k_cr_0)x\frac{\partial x_c(z,t)}{\partial z} \tag{9}$$

Let $x = x_c(z, t)$ in Eq. (9), with Eq. (1), the longitudinal coherent dipole mode space charge field on local centroid becomes:

$$E_s^{(1)}(z,t) = -i\frac{\Lambda_0\hat{a}_c^2}{2\pi\varepsilon_0 r_0}k_c^2K_1(k_cr_0)e^{2i(k_cz-\omega_ct+\phi_c)} = -i\frac{\Lambda_0}{2\pi\varepsilon_0 r_0}k_c^2K_1(k_cr_0)x_c^2(z,t) \tag{10}$$

Eq. (8) and Eq. (10) show the radial and longitudinal coherent dipole mode space charge fields on local centroid are proportional to $x_c$ and $x_c^2$, respectively. This property is determined by the paraxial electric field potential as shown in Eq. (3.33) of [10].

Eqs. (4)(5)(10) give the total longitudinal coherent space charge field on a local beam centroid as:

$$E_s(z,t) = -i\frac{\hat{\Lambda}e^{i(kz-\omega t+\phi)}}{\pi\varepsilon_0 r_0^2 k}[1-kr_0K_1(kr_0)] - i\frac{\Lambda_0\hat{a}_c^2}{2\pi\varepsilon_0 r_0}k_c^2K_1(k_cr_0)e^{2i(k_cz-\omega_ct+\phi_c)} \tag{11}$$

The longitudinal dipole mode space charge wake potential over ring circumference $C_0$ [11, 12] is:

$$V = E_s^{(1)}(z,t)C_0 = -Z_{1,sc}^{\parallel} I_0 x_c^{\,2}(z,t) \tag{12}$$

where $Z_{1,SC}^{\parallel}$ is the longitudinal dipole mode space charge impedance with a unit of $\Omega m^{-2}$, which can be calculated from Eq. (10) and Eq. (12) as:

$$Z_{1,sc}^{\parallel} = i\frac{Z_0 R}{\beta r_0} k_c^{\,2} K_1(k_c r_0) \tag{13}$$

Figure 6 shows the calculated longitudinal monopole and dipole modes space charge impedances of a circular $H_2^+$ beam with emittance of 50 π mm*mrad and kinetic energy of 19.9 keV in SIR.

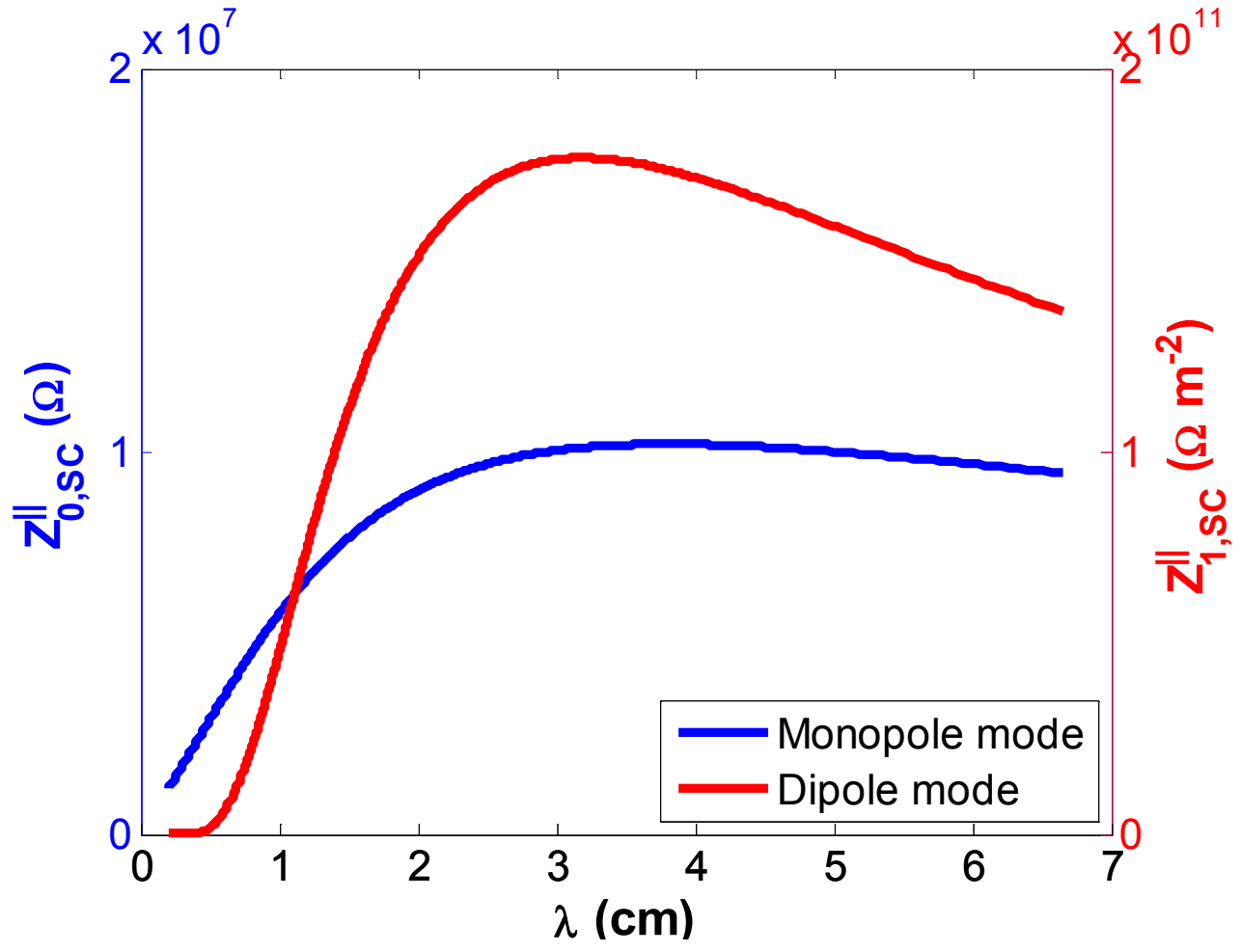

FIG. 6. (Color) Longitudinal monopole and dipole modes

space charge impedances.

## B. Radial and longitudinal equations of motion

According to Eq. (5) of [6], Eq. (1.45) of [11] and Eq. (11), the radial and longitudinal equations of motion of local beam centroid are:

$$x_c^{''} + \frac{\nu_{coh}^2(k_c)}{R^2} x_c = \frac{\delta_c}{R} \tag{14}$$

$$z' = -\eta\delta_c \tag{15}$$

$$\delta_c^{'} = -i\frac{e\hat{\Lambda}e^{i(ks-\omega t+\phi_0)}}{\pi\varepsilon_0 r_0^{\,2} k\beta^2 E}[1 - kr_0 K_1(kr_0)] - i\frac{e\Lambda_0 \hat{a}_c^{\,2}}{2\pi\varepsilon_0 r_0 \beta^2 E} k_c^{\,2} K_1(k_c r_0) e^{2i(k_c s-\omega_c t+\phi_{0,c})} \tag{16}$$

Where $\delta_c$ is the fractional momentum deviation of local beam centroid, $\nu_{coh}(k_c)$ is the coherent radial betatron tune taking to account the coherent linear radial space charge field due to centroid wiggles with modulation wavenumber $k_c$ as expressed in Eq. (7), η is slip factor, e is unit charge, E is total energy of single particle. The prime stands for differentiation with respect to path length s. If $\delta_c$ (s = 0) = 0 for the beam used in the simulations, then $\delta_c$ can be calculated by integration as:

$$\delta_c(s,t) = -\frac{e\hat{\Lambda}e^{i(ks-\omega t+\phi_0)}}{\pi\varepsilon_0 r_0^2 k^2 \beta^2 E}[1-kr_0K_1(kr_0)] - \frac{e\Lambda_0 \hat{a}_c^2}{4\pi\varepsilon_0 r_0 \beta^2 E} k_c K_1(k_c r_0) e^{2i(k_c s-\omega_c t+\phi_{0,c})} \tag{17}$$

Accordingly, Eq. (14) becomes:

$$x_c'' + \frac{\nu_{coh}^2(k_c)}{R^2} x_c = -\frac{e\hat{\Lambda}e^{i(ks-\omega t+\phi_0)}}{\pi\varepsilon_0 r_0^2 k^2 \beta^2 ER}[1-kr_0K_1(kr_0)] - \frac{e\Lambda_0 \hat{a}_c^2}{4\pi\varepsilon_0 r_0 \beta^2 ER} k_c K_1(k_c r_0) e^{2i(k_c s-\omega_c t+\phi_{0,c})} \tag{18}$$

In the first order approximation, if centroid offset amplitude is small, the second term on RHS of Eqs. (17)(18) induced by longitudinal coherent dipole mode space charge field can be neglected, then Eqs. (17)(18) become:

$$\delta_c(s,t) = -\frac{e\hat{\Lambda}e^{i(ks-\omega t+\phi_0)}}{\pi\varepsilon_0 r_0^2 k^2 \beta^2 E}[1-kr_0K_1(kr_0)] \tag{19}$$

$$x_c'' + \frac{\nu_{coh}^2(k_c)}{R^2} x_c = -\frac{e\hat{\Lambda}e^{i(ks-\omega t+\phi_0)}}{\pi\varepsilon_0 r_0^2 k^2 \beta^2 ER}[1-kr_0K_1(kr_0)] \tag{20}$$

We can see Eq. (20) describes a forced harmonic oscillation which results in $k \approx k_c$, $\omega \approx \omega_c$ and $\mathrm{Im}(\omega) \approx \mathrm{Im}(\omega_c)$, because the instability growth rates $1/\tau = \mathrm{Im}(\omega)$, then in the first order approximation, the growth rate spectra of line charge densities and local centroid offsets are approximately the same. For $k \approx k_c$, if the second terms on RHS of Eqs. (11)(17)(18) are comparable to the first terms thus cannot be neglected, then Eqs. (17)(18) will become nonlinear equations and will not be discussed in this paper due to their complexity.

The above analyses and discussions are based on the assumptions of single harmonic component in longitudinal profiles of both centroid offset $x_c(z, t)$ and line charge density $\Lambda(z, t)$. In reality, the distribution functions of the centroid offset $x_c(z, t)$ and line charge density $\Lambda(z, t)$ of a coasting long bunch in SIR have rich spectrum of modulation wavenumbers $k_c$ and k, respectively. For a given wavenumber k of $\Lambda(z, t)$, besides the harmonic component of centroid offset $x_c(z, t)$ with wavenumber $k_c = k$ as we discussed above, another important harmonic component of $x_c(z, t)$ is the one with wavenumber $k_c = k/2$, in this case the second terms on RHS of Eqs. (11)(17)(18) may have perturbation wavenumber $2k_c = k$, they can linearly superimpose on the first terms of the RHS of the above equations and induce betatron oscillations in the spectral evolutions of line charge densities, radial centroid offsets and energy deviations. This special case will be discussed further below.

## C. Betatron oscillations in evolution of beam parameters

Assume a beam of rich spectrum of k in $\Lambda(z, t)$ and $k_c$ in $x_c(z, t)$, respectively. For each given k, if the longitudinal coherent dipole mode space charge field is taken into account, then we may define a time-dependent equivalent longitudinal monopole mode space charge field $[E_s^{(0)}(z,t)]_{eq}$ and associated equivalent longitudinal monopole mode space charge impedance $[Z^{\|}_{0,sc}(k,t)]_{eq}$ by Eq. (21) -Eq. (24) as below:

$$[E_s^{(0)}(z,t)]_{eq} = E_s^{(0)}(z,t) + E_s^{(1)}(z,t) \tag{21}$$

$$-E_s^{(0)}(z,t)C_0 = Z^{\|}_{0,sc}(k)\hat{I}e^{i(kz-\omega t+\phi)} \tag{22}$$

$$-E_s^{(1)}(z,t)C_0 = Z^{\|}_{1,sc}(k)I_0 x_c^2(z,t) = Z^{\|}_{1,sc}(k)I_0 \hat{a}_c^2 e^{2i(k_c z-\omega_c t+\phi_c)} \tag{23}$$

$$-[E_s^{(0)}(z,t)]_{eq}C_0 = [Z^{\|}_{0,sc}(k,t)]_{eq}\hat{I}e^{i(kz-\omega t+\phi)} \tag{24}$$

Eq. (21) - Eq. (24) give:

$$[Z^{\|}_{0,SC}(k,t)]_{eq} = Z^{\|}_{0,sc}(k) + Z^{\|}_{1,sc}(k)\hat{a}_c^2 \frac{I_0}{\hat{I}} e^{\mathrm{Im}(2\omega_c-\omega)t} e^{i[(2k_c-k)z-\mathrm{Re}(2\omega_c-\omega)t+(2\phi_c-\phi)]} \tag{25}$$

In the case of $k_c \neq k/2$, $[Z^{\|}_{0,sc}(k,t)]_{eq}$ will depend on longitudinal coordinate z, the beam dynamics is nonlinear, this complicated case of coupling will not be discussed in this paper.

In the case of $k_c = k/2$, $[Z^{\|}_{0,sc}(k,t)]_{eq}$ will be independent of longitudinal coordinate z, the coupling of longitudinal space charge field induced by line charge density modulation of wavenumber k and centroid offsets modulation of wavenumber $k_c$ is linear. This special case of $k_c = k/2$ is determined by the property that longitudinal coherent dipole mode space charge field of a beam with centroid wiggles is proportional to $x_c^2$. Usually, $\mathrm{Re}(\omega_c) \approx \omega_\beta$, $\mathrm{Re}(\omega) \approx 0$, where $\omega_\beta$ is the angular betatron frequency. Then Eq. (25) can be simplified as:

$$[Z^{\|}_{0,sc}(k,t)]_{eq} = Z^{\|}_{0,sc}(k)[1 + Ae^{B(t)t}e^{-2i\omega_\beta t}] \tag{26}$$

where

$$A = \frac{Z^{\|}_{1,sc}(k)\hat{a}_c^{\,2}}{Z^{\|}_{0,sc}(k)}\frac{I_0}{\hat{I}}e^{i(2\phi_{0,c}-\phi_0)} \tag{27}$$

$$B(t) = \mathrm{Im}[2\omega_c(t) - \omega(t)] \tag{28}$$

Note A is a time-independent complex number, while B (t) is real and function of time t.

According to Eq. (2), the amplitude of perturbed line charge density can be expressed as:

$$|\Lambda_1(t)| = \hat{\Lambda}e^{\mathrm{Im}[\omega(t)]t} = \hat{\Lambda}e^{\frac{t}{\tau(t)}} \tag{29}$$

When energy spread, local centroid offsets are small, modulation wavelengths of line charge densities are greater than beam diameter, the instability growth rates can be estimated by the conventional 1D microwave instability formula [6]:

$$\tau_0^{-1}(k) = \omega_0\sqrt{-i\frac{\eta e I_0 k R Z(k)}{2\pi\beta^2 E}} \tag{30}$$

Where $\omega_0$ is the angular revolution frequency of on-momentum particles, $I_0$ is unperturbed beam intensity. For bunch in SIR, $\eta(k_c)$ is the coherent slip factor induced by the sinusoidal radial centroid offsets with a wavenumber $k_c$, while $Z$ (k) is the longitudinal monopole mode space charge impedance with longitudinal charge density perturbation wavenumber k and can be expressed by Eq. (6). Because $k \approx k_c$ as discussed in Sec. IV.B, we can use the same k in expressions of $\eta(k)$ and Z(k) in Eq. (30) just as treated in [6] (please check Eqs. (2)(12)(13)(14) in [6]).

If $Z(k)$ of Eq. (30) is replaced by $[Z_{0,sc}^{\|}(k,t)]_{eq}$ of Eq. (26), the time-dependent growth rates become:

$$\tau^{-1}(k,t) = \tau_0^{-1}(k)\,\mathrm{Re}\{[1 + Ae^{B(t)t}e^{-2i\omega_\beta t}]^{\frac{1}{2}}\} \tag{31}$$

The amplitude of line charge density perturbation in Eq. (29) becomes:

$$|\Lambda_1(t)| = \hat{\Lambda} e^{\frac{t}{\tau_0(k)}\mathrm{Re}\{[1+Ae^{B(t)t}e^{-2i\omega_\beta t}]^{\frac{1}{2}}\}} \tag{32}$$

If $\hat{a}_c = 0$, then $|A| = 0$, $\tau^{-1}(k,t) = \tau_0^{-1}(k)$, $|\Lambda_1(t)| = \hat{\Lambda} e^{\frac{t}{\tau_0(k)}}$, it is a pure exponential growth as predicted by the conventional 1D microwave instability growth rate formula Eq. (30). But for a coasting long bunch in SIR, since there are always radial centroid offsets induced by longitudinal space charge forces and the associated energy deviations, the condition $\hat{a}_c = 0$ does not hold.

If $\hat{a}_c \neq 0$, $|A| \neq 0$ and if $|Ae^{B(t)t}| << 1$, because $B(t) = \mathrm{Im}[2\omega_c(t) - \omega(t)] \approx \mathrm{Im}(\omega) = \tau_0^{-1}(k)$, Eq. (32) can be approximated as:

$$|\Lambda_1(t)| \approx \hat{\Lambda} e^{\frac{t}{\tau_0}}[1 + \sqrt{|A|}e^{\frac{t}{2\tau_0}}\cos(\omega_\beta t + \frac{\phi_0}{2} - \phi_{0,c})] \tag{33}$$

Eq. (33) is an exponential growth function modulated by betatron frequency $\omega_\beta$, which can explain the betatron oscillations in spectral evolutions of normalized line charge density in Figure 2. For the same reason, the second terms on RHS of Eq. (18) and Eq. (17) induce the betatron oscillations in spectral evolutions of centroid offsets in Figure 3 and energy deviations in Figure 4, respectively. From Eq. (31) and Eq. (27), we can see the instantaneous microwave instability growth rates $\tau^{-1}(k, t)$ at time t depend on the current modulation strength $\hat{I}/I_0$, the centroid offset amplitude $\hat{a}_c$, the phase angles $\phi_{c0}$ and $\phi_0$.

Now we can fit the curves of evolution of line charge densities in Figure 2 to get the simulated instability growth rates. Eq. (33) may be expressed as a general fitting function below:

$$|\Lambda_1(t)| \approx \hat{\Lambda} e^{\frac{t}{\tau_0}} + Pe^{Qt/T_0}\cos(\omega t + \Phi) \tag{34}$$

where $\hat{\Lambda}$, P, Q, ω, Φ, $\tau_0$ are fit coefficients, $T_0$ is the revolution period of $H_2^+$ ion, $t = N_t T_0$, $N_t$ is the turn number, $1/\tau_0$ is just the long term instability growth rates in the first order approximation.

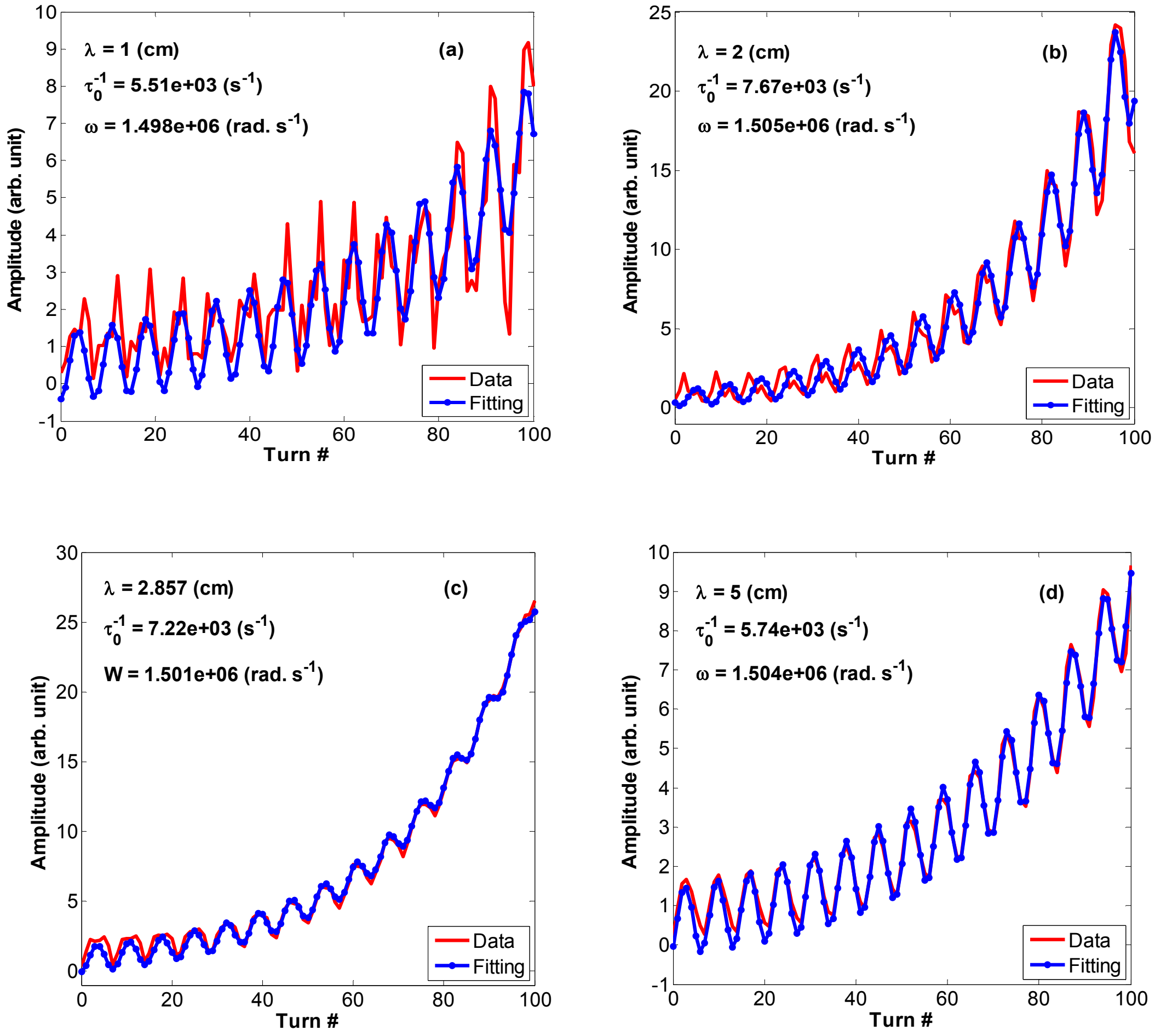

FIG. 7. (Color) Curve fitting results for growth rates of line charge densities

(a). λ = 1.0 cm; (b). λ = 2.0 cm; (c). λ = 2.857 cm; (d). λ = 5.0 cm;

The curve fitting results for growth rates of line charge densities of some chosen modulation wavelengths are show in Figure 7. For beam energy of 19.9 keV, the nominal angular betatron frequency is $\omega_\beta = 1.499 \times 10^6$ radian/second. We can see these curves can mainly be decomposed into exponential growths and betatron oscillations. This result is consistent with the theoretical prediction deduced from the mechanism of longitudinal coherent dipole mode space charge field discussed in this section. Figure 7 also shows that for λ ⩾ 1.0 cm, the linear coupling effects of the centroid offset modulation with wavenumber $k_c = k/2$ (or wavelength $\lambda_c = 2\lambda$) on line charge density modulation with wavenumber k dominate over the other nonlinear coupling effects with wavenumber $k_c \neq k/2$ (or wavelength $\lambda_c \neq 2\lambda$). The special coupling mode between $k_c = k/2$ and k is linear since longitudinal coherent dipole mode space charge field $E_s^{(1)}$ is proportional to $x_c^2$ which has a wavenumber of $2k_c = k$. This field may directly superimpose on and couple with the conventional longitudinal monopole mode space charge field $E_s^{(0)}$ induced by line charge density modulation with wavenumber of k. For all other cases of coupling between k and $k_c \neq k/2$, for example $k_c$ = k, 2k, 3k, 4k…., etc, since $x_c^2$ has a wavenumber of $2k_c \neq k$, then $E_s^{(1)}$ induced by these $k_c$ cannot linearly superimpose on $E_s^{(0)}$ of wavenumber k, the coupling between $E_s^{(0)}$ and $E_s^{(1)}$ for these modes is nonlinear.

Note that if perturbation wavelengths are comparable to or less than the beam radius, $\lambda \leq 0.5$ cm, because of the Landau damping effects due to beam emittance and energy spread, the exponential growths of microwave instability for these short perturbation wavelengths are strongly suppressed, the curves of evolution of line charge densities have more jitters with short perturbation wavelengths, then besides the wavenumber $k_c = k/2$, the nonlinear coupling effects on the field $E_s^{(1)}$ induced by other wavenumbers $k_c$ are strong and must be taken into account too, especially the wavenumber of $k_c = k$, in this case, the Eqs. (31)(33)(34) do not hold.

Now we can see the three components of space charge field in Figure 5 play different roles in microwave instability in isochronous ring: $E_x$ raises the working point above transition and enhances the microwave instability [6], $E_s^{(0)}$ together with $E_x$ mainly induce exponential growth of amplitudes of perturbed line charge densities [6], and $E_s^{(1)}$ with $k_c = k$ mainly induces betatron oscillations in the evolution of perturbed line charge densities, sometimes it enhances the instability growth, sometimes it suppresses the instability growth, it depends on if $E_s^{(0)}$ and $E_s^{(1)}$ are in phase or out of phase. This effect can be observed clearly in Figure 7. If we average the effects of $E_s^{(1)}$ on spectral evolution of line charge densities over a large time scale, we will see usually the time-integrated effects are small compared with the exponential growth components, thus can be neglected in the long-term instability growth.

When the instability growth rates are calculated by the method of FFT and curve fitting, special care should be paid on the second order effects induced by $E_s^{(1)}$, especially when the width of time window used in the curve fitting is shorter than the betatron oscillation period. The conventional 1D growth rate formula Eq. (30) is only valid for prediction of long-term instability growth rates of mono-energetic beams or beams with small energy spread neglecting Landau damping effects and the second order effects due to $E_s^{(1)}$.

In summary, in the first order approximation, the effects of $E_s^{(1)}$ are neglected, the microwave instabilities in isochronous ring demonstrate pure exponential growths as predicted by Eq, (30), because the line charge densities are the driving forces for the resonant growths of modulation amplitudes of centroid offsets and energy deviations as shown in Eq. (20) and Eq. (19), respectively, the modulation amplitudes of centroid offsets and energy deviations have the same growth rates as those of the line charge densities. In the second order approximation, the effects of $E_s^{(1)}$ induced by centroid wiggles of various wavenumber $k_c$ should be taken into account, then the growths of microwave instability are not pure exponential functions of time any more. In this paper, a wavenumber of $k_c = k/2$ is used as a special example of liner coupling, then the microwave instabilities in isochronous ring are characterized by betatron oscillations superimposed on pure exponential growths as predicted by Eq, (31). By this way, we can see how the betatron oscillations and growth rates of modulation amplitudes of line charge densities, centroid offsets and energy deviations of coasting long bunch couple and interact with each other in isochronous regime.

## V. CONCLUSION

In this paper we explored beam dynamics with space charge in isochronous regime. By simulation studies, the betatron oscillations in spectral evolutions of line charge densities, radial centroid offsets and coherent energy deviations are discovered. These phenomena were explained by longitudinal coherent dipole mode space charge field with the concept of time-dependent equivalent longitudinal space charge impedance for the case of $k_c = k/2$. A real coasting long bunch in SIR usually have rich spectrum of modulation wavenumbers $k_c$ of centroid offset $x_c(z, t)$ and k of line charge density $\Lambda(z, t)$, respectively. For each given k, there are two special values of $k_c$: $k_c = k$ and $k_c = k/2$. In the first order approximation, the former case excites the resonant exponential growths of amplitudes of line charge density $\Lambda(z, t)$, centroid offset $x_c(z, t)$, and coherent energy deviation $\Delta E_c(z, t)$. In the second order approximation, the latter case induces the betatron oscillations in spectral evolutions of these parameters. This paper shows that for a long coasting bunch with space charge in isochronous regime, both the radial and longitudinal coherent dipole mode space charge fields should be taken into account in the beam dynamics. The spectra of line charge densities, local centroid offsets and energy deviations may interact with each other and evolve in a self-consistent way.

An accurate prediction of instability growth rates in isochronous regime requires a more complicated 2D dispersion relation incorporating Landau damping effects. This will be discussed in the future works.

## ACKNOWLEDGEMENTS

We would like to thank the guidance of Prof. F. Marti and T. P. Wangler, we are also grateful to Y.C. Wang, S. Y. Lee, K. Y. Ng, G. Stupakov, E. Pozdeyev, A. W. Chao, R. York, M. Syphers, V. Zelevinsky, J. Baldwin, and J. A. Rodriguez for their fruitful discussions and suggestions. This work was supported by NSF Grant # PHY 0606007.

FIG. 1. (Color) Evolution of top views of beam profiles.

FIG. 2. (Color) Evolution of harmonic amplitudes of normalized line charge densities.

FIG. 3. (Color) Evolution of harmonic amplitudes of radial centroid offsets.

FIG. 4. (Color) Evolution of harmonic amplitudes of energy deviations.

FIG. 5. (Color) Space charge components on local centroid with coordinate ($x_c$, z) at time t.

FIG. 6. (Color) Longitudinal monopole and dipole modes space charge impedances.

FIG. 7. (Color) Curve fitting results for growth rates of line charge densities

(a). $\lambda$ = 1.0 cm; (b). $\lambda$ = 2.0 cm; (c). $\lambda$ = 2.857 cm; (d). $\lambda$ = 5.0 cm.